\documentstyle[nato,epsf]{crckapb}

\newcommand{\be}{\begin{equation}}
\newcommand{\ee}{\end{equation}}
\newcommand{\bdm}{\begin{displaymath}}
\newcommand{\edm}{\end{displaymath}}

\def\scro{{\cal O}}

\def\su3{$SU(3)$}

\begin{opening}
\title{First experiences with HMC for dynamical overlap fermions
\vskip-3.7cm\hfill \vbox{\hbox{\rm FSU-SCRI-99C-73} 
\hbox{\rm JLAB-THY-99-40}}\vskip2.3cm
}

\author{Achim Bode}
\author{Urs M. Heller}
\institute{
SCRI, Florida State University, \\
Tallahassee, FL 32306-4130, USA}
\author{Robert G. Edwards}
\institute{
Jefferson Lab,
12000 Jefferson Avenue, \\
MS 12H2,
Newport News, VA 23606, USA}
\author{Rajamani Narayanan}
\institute{
American Physical Society, \\
One Research Road,
Ridge, NY 11961, USA}

\end{opening}

\runningtitle{First experiences $\dots$ overlap fermions}

\begin{document}

\begin{abstract}
We describe\footnote{Poster presented at the workshop ``Lattice fermions
and structure of the vacuum'', October 5--9, 1999, Dubna, Russia.}
an HMC algorithm for dynamical overlap fermions which
makes use of their good chiral properties. We test the algorithm
in the Schwinger model. Topological sectors are readily changed
even in the massless case.
\end{abstract}

\section{HMC algorithm for overlap fermions for any number of flavors}

Overlap fermions represent a lattice discretization of fermions with
the same chiral properties as continuum fermions~\cite{HN1}. Properties
of overlap fermions are reviewed in~\cite{Dubna} (see also \cite{EHN1}).
In this contribution we would like to describe a Hybrid Monte
Carlo (HMC) algorithm for the dynamical simulation of overlap fermions,
which exploits some of their chiral properties.

We denote by $H_o(\mu)$ the hermitian overlap Dirac operator $\gamma_5
D(\mu)$ and find $D^\dagger(\mu) D(\mu) = H^2_o(\mu)$.
Since $[H_o^2(\mu), \gamma_5] = 0$~\cite{Dubna,EHN1} one can split
$H_o^2(\mu)$ into two parts, each acting in one chirality sector only,
$H_o^2(\mu) = H_{o+}^2(\mu) + H_{o-}^2(\mu)$ where, with
$P_\pm = \frac{1}{2} (1 \pm \gamma_5)$,
\be
H_{o\pm}^2(\mu) = \frac{1+\mu^2}{2} P_\pm
 \pm \frac{1-\mu^2}{2} P_\pm \epsilon(H_w) P_\pm .
\label{eq:hsq_proj}
\ee

The non-zero eigenvalues of $H_o^2(\mu)$ are equal in both chirality
sectors and hence also their contribution to the fermion determinant:
\be
\det(H_{o+}^{\prime 2}(\mu)) = \det(H_{o-}^{\prime 2}(\mu)) > 0
\ee
The $^\prime$ indicates that the zero modes have been left out.

For $N_f$ dynamical flavors the fermion determinant is thus
\begin{eqnarray}
[\det(D(\mu))]^{N_f} &=& \mu^{N_f|Q|} [\det(D^\prime(\mu))]^{N_f} =
\nonumber \\
 \mu^{N_f|Q|} [\det(H_o^{\prime 2}(\mu))]^{N_f/2}
 &=& \mu^{N_f|Q|} [\det(H_{o\pm}^{\prime 2}(\mu))]^{N_f} .
\end{eqnarray}
We can use this rewriting to get a Hybrid Monte Carlo
algorithm for dynamical overlap fermions for any number of flavors.
For each flavor we introduce one pseudo-fermion of a {\bf single}
chirality:
\be
\det(H^{\prime 2}_{o\pm}(\mu)) = \int d\phi_\pm^\dagger\phi_\pm
   e^{-S_p}~; \qquad
   S_p = \phi_\pm^\dagger\left [H^{\prime 2}_{o\pm}(\mu)\right]^{-1}\phi_\pm
   ~.
\ee
The choice of the chirality is made such as to avoid zero modes: If the
gauge configuration at the beginning of the trajectory has non-trivial
topology, we choose the chirality that does not have an exact zero
mode of the massless overlap Dirac operator. If the topology is trivial,
we choose the chirality randomly.
To take the zero mode contribution into account, we reweight to compute
observables
\be
\langle\scro\rangle \ = \ {{\langle \mu^{N_f |Q|}\scro\rangle_\pm}\  / \ 
    {\langle\mu^{N_f |Q|}\rangle_\pm}} .
\ee

Having introduced the pseudo-fermions, doing HMC is straightforward.
We need the contribution from the pseudo-fermions to the force:
\be
\frac{\delta S_p}{\delta U} = 
\mp \frac{1}{2} (1-\mu^2) \chi^\dagger_\pm \frac{\delta \epsilon(H_w)}{\delta U}
\chi_\pm~;\qquad
\bigl[H_{o\pm}^2(\mu)\bigr]^{-1} \phi_\pm = \chi_\pm .
\ee
We use a rational polynomial
approximation for $\epsilon(H_w)$ written as a sum over poles~\cite{HN2,EHN2}:
\be
\epsilon(x) \leftarrow x \frac{P(x^2)}{Q(x^2)} = 
  x \Bigl(c_0 + \sum_k \frac{c_k}{x^2 + b_k}\Bigr) .
\ee
Straightforward algebra then gives (see also \cite{Liu})
\begin{eqnarray}
\chi^\dagger_\pm \frac{\delta \epsilon(H_w)}{\delta U} \chi_\pm &\leftarrow&
c_0 \chi^\dagger_\pm \frac{\delta H_w}{\delta U} \chi_\pm +
\sum_k c_k b_k \chi^\dagger_{k\pm} \frac{\delta H_w}{\delta U} \chi_{k\pm}
\nonumber \\
&& - \sum_k c_k \chi^\dagger_{k\pm} H_w \frac{\delta H_w}{\delta U} H_w
\chi_{k\pm} .
\end{eqnarray}
where we introduced
\be
\chi_{k\pm} = [ H^2_w + b^2_k ]^{-1} \chi_\pm .
\ee
The computation of the force requires thus one additional multi-shift
``inner'' CG inversion to obtain the $\chi_{k\pm}$.

A few remarks are in order: (1) We anticipate that a straightforward
HMC for dynamical overlap fermions will suffer even more than with
staggered fermions from difficulties in changing topology due to the
existence of exact zero modes. By working only in one chiral sector,
a change of topology is possible, unimpeded by the fermions, as long as
the number of zero modes changes only in the opposite chirality sector.
(2) Accuracy of the approximation of $\epsilon(H_w)$ can be enforced
by projecting out the lowest few eigenvectors of $H_w$, and adding their
correct contribution exactly~\cite{Dubna}. The molecular dynamics
evolution of the eigenvector projectors $P_\pm$ in Eq. (\ref{eq:hsq_proj}) can
be included using ordinary first order perturbation theory.
However, we have not included projections in our dynamical fermion code yet.
(3) The approximation of $\epsilon(H_w)$ used in the molecular dynamics
steps need not be the same as the approximation of $\epsilon(H_w)$ for the
Metropolis accept/reject step. {\it E.g.} an approximation,
which is smooth around the origin, can be used for the HMD part and the
more accurate optimal rational approximation with projection for the
accept/reject step.

\section{Testing in the Schwinger model}

\begin{figure}
\begin{center}
\vspace*{5mm}
\epsfxsize 80mm
\centerline{\epsfbox[70 70 550 580]{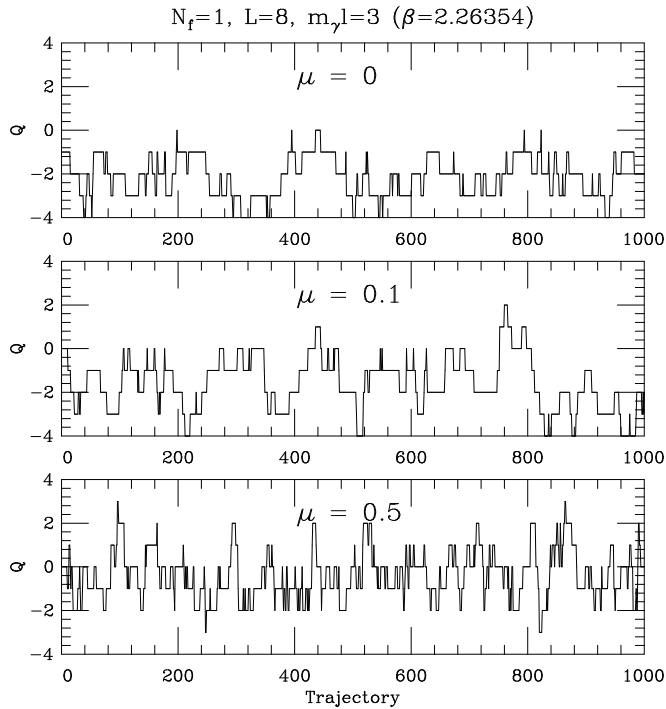}}
\end{center}
\caption{Time history of $Q$ in the 1 flavor model for several quark
masses.}
\label{fig:Q_nf1}
\end{figure}

We tested our HMC algorithm in the $N_f=1$ and 2 Schwinger model.
We first look at time histories of the topological charge $Q$, determined
via the number of exact zero modes. We see (Fig.~\ref{fig:Q_nf1}) that the
topological charge changes, even in the massless case.

We compared our HMC results with fiducial
results, obtained by a brute force approach (exact diagonalization, and
then reweighting with the fermion determinant of quenched gauge fields).
We notice that the acceptance rate does not drop rapidly and the
number of CG iterations does not diverge as $\mu \to 0$.

\section{Conclusions}

The non-zero eigenvalues of $H_o^2(\mu)$ in each chirality
sector contribute identically to the overlap fermion determinant.
Utilizing this fact, and separating the contribution from
the fermion zero modes in non-trivial gauge fields, we devised an
HMC algorithm for any number of flavors of overlap fermions, with
changes of topology possible even in the massless limit.
The trick consists in working in the chirality sector without
exact zero modes.

Preliminary tests in the $N_f=1$ and 2 Schwinger model show
that the algorithm works. The topological charge changes.
The algorithm works even in the massless case. The acceptance
rate does not go to zero or the CG count to infinity.
Further tests on larger systems and in four dimensions are needed
to better judge the usefulness of the algorithm for realistic
dynamical simulations.

This work has been supported in part by DOE contracts DE-FG05-85ER250000 and
DE-FG05-96ER40979. We would like to thanks the organizers for the
opportunity to present this poster during the workshop.


\begin{thebibliography}{99}

\bibitem{HN1} H. Neuberger, {\it Phys. Lett.} {\bf B417} (1998) 141.

\bibitem{Dubna} R.G. Edwards, U.M. Heller. J. Kiskis and R. Narayanan,
these proceedings.

\bibitem{EHN1} R.G. Edwards, U.M. Heller and R. Narayanan,
{\it Phys. Rev.} {\bf D59} (1999) 094510.

\bibitem{HN2} H. Neuberger, {\it Phys. Rev. Lett.} {\bf 81} (1998) 4060.

\bibitem{EHN2} R.G. Edwards, U.M. Heller and R. Narayanan,
{\it Nucl. Phys.} {\bf B540} (1999) 457; \\
{\it Parallel Computing} {\bf  25} (1999) 1395.

\bibitem{Liu} C. Liu,  {\it Nucl. Phys.} {\bf B554} (1999) 313.

\end{thebibliography}
\end{document}